
\hyphenation{di-ver-gences}
\magnification 1200
\raggedbottom
\overfullrule = 0pt
\baselineskip 15 pt

\def\Cbar{{\overline C}}

\def\fcal{{\cal F}}

\def\kmag{\vert{\vec k}\vert}
\def\section#1#2{\medskip\noindent{\bf#1.\quad
#2}\par\smallskip}
\def\abstract#1{\vskip 15 pt\midinsert\narrower
\smallskip
 \noindent{\bf ABSTRACT}\quad{#1}
\smallskip
\endinsert}

\def\cen{\centerline}
\def\ntilde{{\tilde n}}

\def\kvec{{\vec k}{}^2}

\def\k2{k^2}

\def\Pstar{{{~}^*\Pi}}
\def\PPstar{{{}^*\Pi}}
\def\tran{\k2_{{T}}}
\def\long{\kvec_{{L}}}

\def\nn{{\ntilde^{\mu}\ntilde^{\nu}\over \ntilde^2}}

\def\nnff{{\ntilde^{\mu}\ntilde^{\nu}\over \ntilde^4}}
\def\nkup{(\ntilde^{\mu}k^\nu+\ntilde^{\nu}k^\mu)}
\def\nk{(\ntilde_{\mu}k_\nu+\ntilde_{\nu}k_\mu)}
\def\kkup{{k^\mu k^\nu\over\k2}}
\def\kk{{k_\mu k_\nu\over\k2}}
\def\n2{\ntilde^2}
\def\tcal{{\cal T}}
\def\lcal{{\cal L}}

\def\plong{{\vec p}\,^2_L}
\def\qlong{{\vec q}\,^2_L}

\def\gn2{{g^2N\over2}}
\def\softint{Im\int_{soft}dp}

\smallskip
\def\section#1#2{\medskip\noindent{\bf#1.\quad
  #2}\par\smallskip}
\def\cen{\centerline}
\cen{{\bf GAUGE DEPENDENCE OF THE RESUMMED }}

\cen{{\bf THERMAL GLUON SELF
ENERGY}}
\vskip 10 pt
\vskip 20 pt
\centerline
{\bf R.~Baier$^1$, G.~Kunstatter$^2$ and D.~Schiff$^3$}
\vskip 10 pt
\centerline{{\it $^{1 }$Fakult\"at f\"ur Physik,
Universit\"at Bielefeld, D-4800 Bielefeld 1, Germany}}
\centerline{{\it $^{2 }$Winnipeg Institute for Theoretical
Physics and Physics Department,}}
\centerline{{\it University of
Winnipeg, Winnipeg, Manitoba, Canada R3B~2E9}}
\centerline{{\it $^{3 }$LPTHE{$^{\dagger}$}, Universit\'e
Paris-Sud, B\^atiment 211, F-91405 Orsay, France}}
\vskip 10pt
\noindent
\footnote{} {$^{\dagger}$ Laboratoire associ\'e du
Centre National
de la Recherche Scientifique}
\centerline{PACS number:~12.38 Mh 12.38 Cy}
\def\ref#1{[#1]}
\vskip 25 pt
\noindent
\abstract{The gauge dependence of the hot gluon self energy is
examined in the context of Pisarski's method for
resumming hard thermal loops. Braaten and Pisarski have used the
Ward identities satisfied by the hard corrections to the n-point
functions to argue the gauge fixing independence of the leading
order resummed QCD plasma
damping rate in covariant and strict Coulomb gauges. We
extend their analysis to include all linear gauges that preserve
rotational invariance and display explicitly the conditions
required for gauge fixing independence. It is shown that in
covariant gauges the resummed damping constant is
gauge fixing independent only if an infrared regulator is
explicitly maintained
throughout the calculation.}
\vskip 40 pt
\noindent
BI-TP 92/19
\par\noindent
LPTHE-Orsay 92/32
\par\noindent
WIN-TH-92/02
\par\noindent
June 1992
\vskip 20 pt
\eject
\section{1}{INTRODUCTION}
There has recently been  a great deal of interest in  QCD at high
temperature, in part due to the hope that a
new state of matter, the quark gluon plasma, might soon be
observed in relativistic heavy ion collisions\ref{1}. There are,
however, very few observable properties of this plasma that can
be predicted directly from the fundamental theory. Most
experimental properties are derived from
phenomenological models that are often {\it ad hoc} and  contain
many
parameters. The few physical quantities
that can be calculated using perturbative QCD have recently been
the subject of much controversy\ref{2}, in part due to the
apparent
gauge fixing dependence of the gluon plasma damping constant at
one
loop[3]. This gauge
fixing dependence has been attributed by some to
the supposed
breakdown of  perturbative QCD\ref{4}, and by others to the need
for a manifestly gauge covariant or invariant linear response
theory\ref{5}.
The correct resolution to this puzzle has recently been proven
to lie in the innacuracy of the naive loop expansion as a
self-consistent approximation scheme: an
accurate, and gauge independent damping constant can only be
obtained by  resumming an infinite number
of loop diagrams.  Although the need to re-sum was realized
fairly early\ref{6}, an explicit prescription for performing the
necessary resummation has only recently been provided by
Pisarski\ref{7}. The proof that the resummation does indeed yield
a gauge independent damping rate was provided using two
distinct methods. The first, due to Braaten and
Pisarski (BP)\ref{8}, involved explicitly performing the
resummation required and using  Ward identitities satisfied by
the improved  perturbative n-point functions to show that the
same resummed damping constant is obtained to leading order in
strict Coulomb gauge and in all covariant gauges. A similar
result was also obtained in covariant gauges using a partial
resummation technique\ref{9}. The second method, due to
Kobes, Kunstatter and Rebhan (KKR)\ref{10} used properties of the
generating functional for QCD to provide a non-perturbative proof
of  the gauge fixing independence of the locations of the
physical poles of  the gluon propagator.
\par
In effect, both the BP and KKR proofs were based on the fact that
the gauge
dependence of the physical structure functions vanishes ``on-
shell". That is, given the structure function $f(k)$ whose zeroes
determine the location of the physical poles in the gluon
propagator, one can show that
$$
\Delta f(k) = f(k) \Delta X(k)~,
\eqno(1.1)
$$
where $\Delta X(k)$ can be calculated perturbatively. Eq.(1.1)
has the immediate consequence that the physical pole position
(and hence also the mass and damping rates, which are essentially
the real and imaginary parts of the pole) are gauge fixing
independent, providing that  $\Delta X(k)$ does not have poles
when $f(k)=0$. KKR provided a non-perturbative proof of
(1.1) for gauges theories in general, and then discussed how to
apply the result to a self-consistent approximation for the gluon
plasma damping rate.  BP on the other
hand, provided a
perturbative proof of (1.1) using the resummed Feynman rules
appropriate for finite temperature QCD.  Clearly, the  former
proof was very
general and necessarily quite formal, while the latter proof
considered only a  restricted family of gauges. Moreover
the proof of Braaten and Pisarski
was also formal, to the extent that it did  not explicitly
calculate the particular $\Delta X(k)$ that arose,  but instead
assumed that it would be well behaved. KKR argued that the pole
structure of
$\Delta X(k)$ would be essentially determined by the poles of the
full ghost propagator, and hence distinct in general from the
solutions to the physical dispersion relations, but as will be
shown below,
their analysis must be re-evaluated in the presence of infrared
divergences.
\par
The purpose of the present paper is to extend the Ward identity
analysis of Braaten and Pisarski to display explicitly the gauge
fixing dependence of all the components of the resummed gluon
self energy and to include all possible linear gauges in which
rotational symmetry (in the rest frame of the heat bath) is
unbroken. Thus we consider
all covariant and non-
covariant gauges in which the four-velocity of the heat bath is
the only additional vector that appears.
\par
The paper is organized as follows:  In Section 2,
we present the Feynman rules for one loop QCD calculations in the
class of gauges considered. This serves to establish conventions
and notation and also to remind the reader of the source of the
tree level Ward identities obeyed by the bare n-point functions.
 Section 3 explains briefly the need for considering higher loop
diagrams, and describes how a corrected damping rate can be
obtained using the
resummation technique proposed by Pisarski\ref{7} and
implemented
by
Braaten and Pisarski\ref{8}.  Section 4 uses the
Ward identities to analyze in detail the gauge dependence of the
gluon self energy, and its associated damping rate,
in this improved approximation scheme. In Section 5 we examine
the integrals that must be
evaluated in Covariant and Coulomb gauges in order to prove gauge
fixing independence of the damping constant and find the
surprising result that in covariant gauges, the imaginary part
of
$\Delta X(k)$ does develop poles on the physical mass shell and
leading to
gauge dependent damping rates unless an infrared regulator is
maintained throughout the calculation. This extends to  gluons
a similar result concerning the gauge dependence of the
resummed fermion damping rate\ref{11}. Section 6 closes with
conclusions.
\vskip 20 pt
\vfill\eject
\section{2}{PRELIMINARIES}
For simplicity we consider only gluons. The discussion can be
generalized to include quarks without much difficulty.
The Feynman rules for the theory are
derivable
from a generating functional of the form:
$$
Z=\int \prod_{a,\mu,x}dA^a_\mu(x)d\Cbar^a(x) dC^a(x) \exp
(I_{cl} +
I_{g.f.} + I_{gh})~ .
\eqno(2.1)
$$
In the above, $I_{cl}$ is the classical $SU(N)$ Yang-Mills
action:
$$
I_{cl}=-{1\over4}\int d^4x F^{a\,\mu\nu}F^a_{\mu\nu}  ~,
\eqno(2.2)
$$
where the colour indices ${a,b,c..}$ run from 1 to $N^2-1$ and
the Yang-Mills field strength is
$$
F^a_{\mu\nu}=\partial_\mu A^a_\nu-\partial_\nu A^a_\mu
   -gf^a_{bc}A^b_\mu A^c_\nu ~.
\eqno(2.3)
$$
The classical action is invariant under the following
infinitesmal gauge
transformations of the vector potential:
$$
\delta A^a_\mu(x)= D_{\mu}^{ac}(x)\lambda^c(x) ~ ,
\eqno(2.4)
$$
where we have defined the covariant derivative operator:
$$
D^{ab}_\mu(x)\equiv \delta^{ab}\partial_\mu-gf^{abc}A^c_\mu(x)
\eqno(2.5)
$$
and $\lambda^a(x)$ is an arbitrary, infinitesmal Lie algebra
valued field.
The fields $\Cbar^a$ and $C^a$ are the Grassman-valued
Faddeev-Popov ghosts necessary to make the source-free generating
functional
independent of the  gauge fixing term $I_{g.f.}$, which is of the
form:
$$
I_{g.f.}= -{1\over2\xi}\int d^4x (\fcal^a)^2    ~ ,
\eqno(2.6)
$$
where $\fcal^a$ is, in principle, an arbitrary function
of the vector potential, restricted only by the condition that
its associated Faddeev-Popov operator be invertible. We will
restrict the consideration
to linear gauges
$$
\fcal^a=\fcal^\mu(x) A^a_\mu(x)~ ,
\eqno(2.7)
$$
where $\fcal^\mu$ is an arbitrary (possibly non-local)
differential operator, in which case the Faddeev-Popov operator
takes the form:
$$
Q^{ab}\equiv {\delta \fcal^b\over \delta A^c_\mu(x)}
D^{ac}_\mu(x)
={\cal F}^\mu (x)D^{ab}_\mu(x)  ~ .
\eqno(2.8)
$$
The ghost action is then :
$$
I_{gh}= \int d^4x \Cbar^a \fcal^\mu D^{ab}_\mu C^b ~.
\eqno(2.9)
$$
\par
Since we are interested in finite temperature Green's
functions\ref{12} using the imaginary time formalism, we will
assume that
we have already Wick rotated to Euclidean space-time, with metric
signature $(++++)$. All functions are therefore periodic in
imaginary time with period
$\beta={1/T}$, as required by the Matsubara formalism. This in
turn requires the
energy, $k_0$, to be an integer multiple of $2\pi T$. The
retarded real time thermal Green's
functions from which the plasma parameters can be extracted are
obtained from the imaginary time Green's functions by
analytically continuing the external
momenta back to real time, after all Euclidean sums and integrals
over internal momenta have been performed.
\par
At finite temperature, the velocity of the heat bath introduces a
preferred direction in space-time which breaks manifest
Lorentz invariance. We will denote this velocity by the four
vector $n^\mu$, and assume that we are in the rest frame of the
heat bath, so that $n^\mu=\delta_0^\mu$. If the gauge fixing
condition does not break rotational invariance, then the momentum
space representation for the most general $\fcal^\mu$ is:
$$
\fcal^\mu=d(k)k^\mu+b(k)n^\mu
\eqno(2.10)
$$
for arbitrary functions $d(k)$ and $b(k)$ of the four momentum
$k^\mu$.
\par
In order to simplify the following discussion it is useful to
introduce some relevant tensors. First of all, we will need the
projection operator:
$$
P^\nu_\mu=\delta^\nu_\mu- {k^\nu k_\mu\over \k2}   ~ ,
\eqno(2.11)
$$
where $\k2\equiv (k^0)^2+\kvec$.
This can be used to isolate the component of $n^\mu$ that is
orthogonal to the four momentum, $k^\mu$, namely:
$$
\ntilde^\mu(k)\equiv P^\mu_\nu n^\nu=\delta^\mu_0-k^\mu {n\cdot
k\over \k2}  ~     .
\eqno(2.12)
$$
Note that $\ntilde^2=\kvec/\k2$.
It will be useful to re-express the momentum space gauge fixing
condition in terms of the resulting orthogonal basis:
$$
\fcal^\mu= a(k)k^\mu+b(k)\ntilde^\mu      ,
\eqno(2.13)
$$
where $a(k)=d(k)+b(k)n\cdot k/\k2$.
\par
Finally, we define:
$$\eqalignno{
A^{\mu\nu}&\equiv P^{\mu\nu}-\nn \cr
&=\pmatrix{0&0\cr
           0&\delta_{ij}-{k_ik_j\over\kvec}}   ~,
&  (2.14)\cr}
$$
which projects out spatially transverse modes, and is
orthogonal to both $k^\mu$ and $\ntilde^\mu$.
\par
The corresponding Feynman rules are as follows:
\par\noindent
The bare 2-point function is
$$
\left(D^{(0)}{}^{-1}\right){}^{ab}_{\mu\nu}=\delta^{ab}\left(
 k^2P_{\mu\nu}+{1\over\xi}(b\ntilde_\mu+ak_\mu) (b\ntilde_\nu +
a k_\nu)\right)  ,
\eqno(2.15)
$$
with associated propagator:
$$
\delta^{ab}D^{(0)}_{\mu\nu}(k)=
\delta^{ab}\left({1\over k^2}P_{\mu\nu}
    -{b\over ak^4}\nk+
    \left({\xi\k2+\n2 b^2\over a^2k^4}\right)\kk
  \right)         .
\eqno(2.16)
$$
The ghost propagator and ghost-ghost-gluon vertex in this family
of gauges are, respectively:
$$
D^{ab}=\delta^{ab}{1\over a(k)\k2}  ~ ,
\eqno(2.17)
$$
and
$$gf^{abc}\Gamma^\mu(k) =gf^{abc}(a k^\mu + b \ntilde^\mu) ~.
\eqno(2.18)$$
Due to the linearity of the gauge fixing condition, the three and
four point functions depend only on the classical action, and
are, respectively :
$$
-igf^{abc}\Gamma^{\mu\nu\rho} (p,q,k) =-igf^{abc}
((p-q)^\rho\delta^{\mu\nu}+\hbox{perm})  ~,
\eqno(2.19)
$$
and
$$
-g^2N\Gamma^{\mu\nu\rho\tau} (p,q,r,s)=
   -g^2N(2\delta^{\mu\nu} \delta^{\rho\tau}-
\delta^{\mu\rho} \delta^{\nu\tau}-\delta^{\mu\tau}
\delta^{\nu\rho})                          ~      ,
\eqno(2.20)
$$
where $p+q+k=0$ and we have traced over the colour indices in
the four point function.
\par
An important aspect of the above bare n-point functions is that
they obey the following Ward identities (suppressing colour
indices):
$$\eqalignno{
k_\mu\Gamma^{\mu\nu}(k)&=0 ~,&  (2.21a)\cr
k_\rho\Gamma^{\mu\nu\rho}(p,q,k)&=-\Gamma^{\mu\nu}(p)
+\Gamma^{\mu\nu}(q) ~,
&(2.21b)\cr
s_\sigma\Gamma^{\mu\nu\lambda\sigma}(p,q,r,s) &=
\Gamma^{\mu\nu\lambda}(p+s,q,r)-\Gamma^{\mu\nu\lambda}(p,q+s,r) ~,
& (2.21c)\cr}$$
where $\Gamma^{\mu\nu}\equiv k^2 P^{\mu\nu}$
refers to the gauge independent
part of the tree level 2-point function in (2.15) above. These
identities follow
directly from the gauge invariance of the classical action from
which the n-point functions were derived.
In particular, one can express the gauge invariance of the
classical action in terms of
the following functional differential equation:
$$
D^{ab}_\mu{\delta I_{cl}\over\delta A^b_\mu(x)}=0  ~,
\eqno(2.22)
$$
which expresses the fact that the directional
derivatives of the action along the gauge orbits are zero. The
Ward
identities then can be derived by functionally differentiating
this equation
with respect to the vector potential, and then evaluating the
resulting expression at $A^a_\mu=0$. Differentiating once yields
(2.21a), twice yields (2.21b) and three times yields (2.21c).
\eject
\smallskip
\section{3}{RESUMMED PLASMA PARAMETERS}
The loop expansion does not provide a
self-consistent approximation for the finite temperature QCD
damping rate. In general, an infinite number of loop orders
contribute to the leading order result\ref{6}. This can be
understood by noting that at finite
temperature,
QCD
contains two independent mass scales: $T$ and $gT<<T$. Momenta
that are of order $T$ or greater are called ``hard", while those
that are of order $gT$ or smaller are called ``soft".
For n-point processes in which all
legs are of order $gT$, one loop corrections in which
all internal momenta are hard contribute to the process to the
same order as tree level[7].
Moreover, these are the only such large
``corrections".
 This implies, for example, that the dispersion relations
can be consistently calculated to  order  $g^2 T^2$
by considering only one loop self energy diagrams with hard
internal momenta. No higher loop corrections are required.
Explicit calculations show that the hard loop self energy diagram
is real. The damping constant therefore appears
only at the next order in the resummed perturbative expansion,
namely $g^2T$. One loop calculations to this order
require
summing over diagrams with soft internal legs and as mentioned
above, for soft processes (internal or external) one
needs to encorporate the hard thermal loop corrections to  the
vertices and the propagator.
\par
It is worth
stressing that this algorithm is only valid for the
leading order ($g^2T$) damping constant, and must
be modified in order to calculate corrections either to the
dispersion relations, or the damping rate. For example, in order
to calculate the real part of the dispersion
relations to order $g^2T$, it is necessary to consider
higher loop diagrams involving the modified n-point
functions[7].
\par
Remarkably, the
hard thermal loop corrections to the n-point functions are gauge
fixing independent and obey the
same Ward identities as the tree-level n-point functions
\ref{7,8,13}. \footnote{$^1$}{The gauge invariance of the
hard thermal loop contributions to the quark self energy was
first shown by Klimov and Weldon\ref{14}.}
Let $\delta\Pi^{\mu\nu}$, $\delta\Gamma^{\mu\nu\rho}$,
and $\delta\Gamma^{\mu\nu\rho\sigma}$ denote the hard thermal
loop corrections to the two, three and four-point functions,
respectively. Then it can be shown that[8,13]:
$$\eqalignno{
k_\mu{}^*\Gamma^{\mu\nu}(k)&=0 ~,&  (3.1a)\cr
k_\rho{}^*\Gamma^{\mu\nu\rho}(p,q,k)&=-{}^*\Gamma^{\mu\nu}(p)
+{}^*\Gamma^{\mu\nu}(q) ~,
&  (3.1b)\cr
s_\sigma{}^*\Gamma^{\mu\nu\lambda\sigma}(p,q,r,s) &=
{}^*\Gamma^{\mu\nu\lambda}(p+s,q,r)
-{}^*\Gamma^{\mu\nu\lambda}(p,q+s,r)  ~,
&  (3.1c)\cr}$$
where we have defined the corrected n-point functions:
$$
{}^*\Gamma^{\mu\nu}=k^2 P^{\mu\nu}+\delta\Pi^{\mu\nu} ~,
\eqno(3.2a)
$$
$$
{}^*\Gamma^{\mu\nu\rho}=\Gamma^{\mu\nu\rho}+
\delta\Gamma^{\mu\nu\rho} ~,
\eqno(3.2b)
$$
$$
{}^*\Gamma^{\mu\nu\rho\sigma}=\Gamma^{\mu\nu\rho\sigma}+
\delta\Gamma^{\mu\nu\rho\sigma} ~      .
\eqno(3.2c)
$$
\par
The tree
level Ward identities obeyed by the ${~}^*\Gamma$ imply that the
resummed n-point functions can be derived from a gauge
invariant (resummed) effective Lagrangian. The existence of this
effective Lagrangian, which has explicitly been constructed in
Refs.\ref{13,15}, plays an important role in understanding the
connection between the BP and KKR proofs of gauge fixing
independence.
\par
In terms of the ${~}^*\Gamma$ n-point functions, the resummed
expression for the leading order
imaginary part of the real time gluon self energy is:
$$\eqalignno{
Im~{{}^*\Pi}^{\mu\nu}(K_0,\kmag)
&={g^2N\over2}Im\int_{soft}
dp {~}^*D_{\alpha\beta}
{~}^*\Gamma^{\mu\nu\alpha\beta}(k,-k,p,-p)\cr
&-{g^2N\over2}Im\int_{soft}dp
{~}^*\Gamma^{\alpha\beta\mu}(p,q,k)
{~}^*D_{\alpha\alpha'}(p){~}^*D_{\beta\beta'}(q)
{~}^*\Gamma^{\alpha'\beta'\nu}(p,q,k)\cr
&-{g^2N}
Im\int_{soft}dp~{(a(p)p^\mu+b(p)\ntilde^\mu)
  (a(q)q^\nu+b(q)\ntilde^\nu(q))\over
  a(p)p^2a(q)q^2} ~,
&  (3.3) \cr}
$$
where ${~}^*D^{\mu\nu}$ is the effective propagator defined as:
$$\eqalignno{
{}^*D^{\mu\nu} &= \left( (D^{(0)}){}^{-1}_{\mu\nu}
+\delta\Pi_{\mu\nu} \right)^{-1}\cr
&=({~}^*\Gamma_{\mu\nu}+ {1\over\xi}\left(b\ntilde_\mu+ak_\mu)
(b\ntilde_\nu+ak_\nu)\right)^{-1}  ~.
&   (3.4)\cr}
$$
In the above and what follows, $Im$ refers to the
imaginary part of the
integrals after analytic continuation of
the external momenta:
$ik_0\to K_0+i\epsilon$, and
$$
\int_{soft} dp  \equiv T\sum_{p_0=2\pi nT}\int_{soft} {d^3 \vec
p\over
(2\pi)^3}   ~ ,
\eqno(3.5)
$$
where the spatial integration  is performed over soft momenta
only, and
therefore must be cut-off at some value $\Lambda$ such that
$gT<<\Lambda<<T$. In the calculation of
the imaginary part of the self energy, which is all that we will
be concerned with here, the precise value of the cut-off turns
out to be irrelevant.
\par
Standard linear response theory\ref{16} implies that the response
of the plasma to small external perturbations is
determined by the physical poles in the retarded thermal gluon
propagator.
It is convenient to write the full propagator in terms of the
hard thermal
loop propagator, ${~}^*D_{\mu\nu}$, (suppressing colour
indices) and the resummed self
energy ${~}^*\Pi_{\mu\nu}$ (with the imaginary part at order
$g^2T$
given in
(3.3)):
\par
$$\eqalignno{
D^{\mu\nu}&=
\left({~}^*D^{-1}_{\mu\nu} +{{~}^*\Pi}_{\mu\nu}
\right)^{-1}\cr
&={1\over \tcal(k)} A^{\mu\nu} +
{1\over\lcal(k)}\left[\ntilde^\mu\ntilde^\nu +
{1\over C(k)}\nkup + {1\over D(k)}\kkup\right] ~.
&    (3.6)\cr}
$$
For the class of gauges given in Section 2, the most general form
of the gluon self energy in momentum space is :
$$
{{}^*\Pi}^{\mu\nu}={{}^*\Pi}_t(k)A^{\mu\nu}+{{}^*\Pi}_l(k)
{\ntilde^\mu\ntilde^\nu\over\ntilde^4}
+{{}^*\Pi}_c(k)\nkup+
{{}^*\Pi}_d(k)\kkup ~.
\eqno(3.7)
$$
Note that it is not transverse in general. The components of the
propagator in (3.6) can be related to those of the self energy as
follows\ref{17}:
$$\eqalignno{
\tcal(k)&\equiv\tran+{{}^*\Pi}_t ~, &(3.8a) \cr
\lcal(k)&\equiv\long+{{}^*\Pi}_l +{b^2\ntilde^4\over\xi} \left[1-
{(1+\xi{{~}^*\Pi}_c/ab)^2\over(1+\xi{{~}^*\Pi}_d/a^2\k2)}\right] ~,
&(3.8b)\cr
C(k)&\equiv -{a\over
b\n2}{(\k2+\xi{{~}^*\Pi}_d/a^2)\over(1+\xi{{~}^*\Pi}_c/ab)} ~,
&(3.8c)
\cr
D(k)&= + {a^2\k2\over\xi}
\left({1+\xi{{~}^*\Pi}_d/\k2 a^2\over\long+{{~}^*\Pi}_l+\ntilde^4
b^2/\xi}\right)     ~,
&   (3.8d)\cr}$$
where we have defined
$$
\tran\equiv \k2+\delta\Pi_T ~,
\eqno(3.9)
$$
and
$$
\long\equiv\kvec+\delta\Pi_L ~ .
\eqno(3.10)
$$
In terms of the above decomposition:
$$
{}^*\Gamma^{\mu\nu}=\tran A^{\mu\nu} +\long \nnff   ~,
\eqno(3.11)
$$
consistent with the Ward identity (3.1a).
\par
The explicit expressions for the hard thermal loop corrections to
the two-point function are, after analytic continutation back to
real time $ik_0\to K_0+i\epsilon\,$\ref{8,14}:
$$
\delta\Pi_T(K_0, \kmag)\sim +{3 m^2_g K_0{}^2\over 2
\kvec}
\left[1-\left(1-{\kvec\over K_0^2}\right){K_0\over2\kmag}
ln
\left({K_0+\kmag\over K_0-\kmag}\right)\right]        ~,
\eqno(3.12)
$$
and
$$
\delta\Pi_L(K_0,\kvec)\sim+3m_g^2 \left[1-
{K_0\over2\kmag}ln\left({K_0+\kmag\over K_0-\kmag}\right)
\right]                                              ~,
\eqno(3.13)
$$
where $m_g^2\equiv g^2N T^2/9$ in the absence of quark fields.
The hard thermal loop corrections to the three and four-point
functions are not required in the
following discussion. In fact, as shown by Braaten and
Pisarski[8],
and elaborated below, the
specific form of the corrected n-point functions is irrelevant to
the formal proof of gauge independence of the resummed plasma
damping rates. It is necessary and sufficient that the n-point
functions
satisfy the  Ward identities (3.1a)-(3.1c).
\par
The physical modes of the gluon in the plasma are transverse with
respect to
the momentum four-vector. As can be seen above, the presence of
the preferred four-vector $\ntilde^\mu$ leads to the existence of
two
independent physical  modes at finite
temperature . The first, determined by the coefficient of
$A^{\mu\nu}$ is the usual spatially transverse gluon mode. The
second is a spatially
longitudinal collective mode which is absent at zero temperature
and is usually called the``plasmon mode".  The
dispersion relations and damping rates for
the gluon and  plasmon modes are therefore obtained from the real
and imaginary parts, respectively of the solutions to:
$$\eqalignno{
\tcal(k)&=0 ~,&(3.14a)\cr
\lcal(k)&=0 ~.&  (3.14b)
\cr}$$
\par
Note that in principle, the non-transverse parts of the
self energy
can contribute to the longitudinal dispersion relation and
damping rate.
However, in linear gauges, ${{~}^*\Pi}_c$ and ${{~}^*\Pi}_d$ are
not independent, but are related
by the following (exact) Ward identity\ref{18}:
$$
\fcal_\mu\fcal_\nu D^{\mu\nu}=\xi  ~ ,
\eqno(3.15)
$$
which, in terms of the parametrizations given above reduces to:
$$
{{}^*\Pi}_d(\long+{{}^*\Pi}_l)=\ntilde^4 k^2{{~}^*\Pi}_c^2  ~.
\eqno(3.16)
$$
Eq.(3.16) can be used to express the longitudinal
structure
function in the following form\ref{17}:
$$
\lcal(k)={(\long+{{~}^*\Pi}_l-{\ntilde^4 b\over a}{{~}^*\Pi}_c)^2
\over
(\long+{{~}^*\Pi}_l+{\xi\ntilde^4\over a^2}{{~}^*\Pi}_c^2)}     ~.
\eqno(3.17)
$$
Thus, in general, it appears that ${{~}^*\Pi}_c$  can not only
shift the
location of the longitudinal pole, but also change
the order of the pole from first order to second order. One of
the results of the present paper, however,
is to show that in linear gauges,
${{~}^*\Pi}_c $ does not in fact contribute to the leading order
dispersion relations. This has previously been only verified in
strict Coulomb and covariant gauges\ref{8}.
\par
Given the two structure functions $\tcal$ and $\lcal$, it is
straightforward to derive the dispersion relations
$K^{t,l}_0=\omega^{t,l}(\vec k)$, which are determined by the
zeros of the structure functions. Normally one assumes that the
system is underdamped, so that the imaginary part of the pole is
small compared to the real part. That is:
$$K_0=\omega_p-i\gamma  ~ ,$$
where $\gamma<<\omega_p$. In this case, given a general structure
function of the form:
$K_0^2-f(K_0,\vec k)$, the dispersion relation
becomes:
$$
\omega_p^2=Re f(\omega_p,\vec k) ~  ,
\eqno(3.18)
$$
while the damping rate is:
$$
\gamma=\left.\displaystyle{{Im f(\omega,\vec k)\over
   \omega^2{\partial\over\partial\omega} \left[{Re f(\omega,\vec
k)\over \omega^2}\right]}}\right|_{\omega=\omega_p} ~.
\eqno(3.19)
$$
\par
For the gluon excitation at rest ($\kmag=0$), Eqs.(3.12) and (3.13)
yield the same mass for the longitudinal
and transverse modes, namely
$$
m_{T,L}^2=m_g^2   ~.
\eqno(3.20)
$$
Application of the above techniques in Coulomb gauge yield a
damping rate of\ref{8}:
$$
\gamma_{T,L}\sim6.63\,{g^2NT\over24\pi} ~.
\eqno(3.21)
$$
In reference [8] the Ward identities were used to prove that the
same damping rate would result in covariant gauges, while
Ref.[10] provided general arguments concerning the gauge
independence of $\gamma$. In the following section we examine
explicitly the gauge dependence of the imaginary part of the
leading order resummed self energy
(3.3) in a general class of linear gauges.
\eject
\section{4}{GAUGE DEPENDENCE OF THE RESUMMED SELF ENERGY}
We will now calculate the gauge dependent part of the imaginary
part
of the resummed
self energy
as given by Eq.(3.3) above.
For convenience we write the propagator (3.4) in the following
form:
$$
{}^*D^{\mu\nu}={1\over{\tran}}A^{\mu\nu}+{1\over\long}\ntilde^\mu
\ntilde^\nu   + {\beta(k)\over\long} \nkup +{\alpha(k)\over
\long}
\kkup  ~,
\eqno(4.1)
$$
where we have defined
$$
\beta(k)\equiv -{b\over a}{\ntilde^2\over k^2}   ~,
\eqno(4.2)
$$
$$
\alpha(k)\equiv {\xi\long\over a^2\k2}+ \k2\beta^2 ~.
\eqno(4.3)
$$
Since the modified 3- and 4-point functions are gauge fixing
independent, all the gauge dependence in the self energy will
come from the gauge dependent terms in the propagator. This can
be isolated by defining:
$$
{}^*D^{\mu\nu}={}^*D'^{\mu\nu}+\Delta {}^*D^{\mu\nu}   ~,
\eqno(4.4)
$$
where we consider the gauge variation about an arbitrary
``fiducial" gauge with resummed propagator:
$$
{}^*D'^{\mu\nu}\equiv{1\over\tran}A^{\mu\nu} + {1\over\long}
\ntilde^\mu\ntilde^\nu
  +{\beta_0(k)\over\long}\nkup +{\beta_0^2\over \long} k^\mu
k^\nu                                           ~.
\eqno(4.5)
$$
Note that the above form of the propagator assumes that $\xi_0=0$
in the fiducial gauge, so that the gauge fixing condition is
enforced by a delta function in the path integral.
For example, if the fiducial gauge is strict Coulomb gauge, then
$\beta_0=k_0/\k2$, while in strict Covariant gauge, $\beta_0=0$.
These examples will be treated in
detail below.
The gauge dependent part or the propagator can now be written:
$$
\Delta {}^*D^{\mu\nu}={\Delta\beta\over\long}\nkup +
{\Delta\alpha\over\long}\kkup                    ~,
\eqno(4.6)
$$
where $\Delta\beta\equiv\beta-\beta_0$ and
$$
\Delta\alpha\equiv \k2(\beta^2 - \beta_0^2)
+{\xi\long\over a^2\k2}                    ~.
\eqno(4.7)
$$
The expression for the resummed self energy $\Pstar^{\mu\nu}$
in
Eq.(3.3) can easily be separated into its gauge independent (for
a specific fiducial gauge) and gauge dependent parts. We consider
each term in expression (3.3) in turn.
The first term, involving the 4-point function, has the following
gauge dependent contribution:
$$\eqalignno{
Im<\Delta {}^*D {~}^*\Gamma^{(4)}>&=
\gn2\softint ~ \Delta
{}^*D_{\alpha\beta} {~}^*\Gamma^{\mu\nu\alpha\beta}\cr
&=\gn2\softint \left[ {\Delta\beta(p)\over\plong}
(\ntilde_\alpha(p)
   p_\beta+\ntilde_\beta(p) p_\alpha) +\right.\cr
 &\qquad\left. {\Delta\alpha\over\plong}
   {p_\alpha p_\beta\over p^2}\right]
{~}^*\Gamma^{\mu\nu\alpha\beta}
   (k,-k,p,-p)\cr
&=g^2N\softint ~ {\Delta\beta(p)\over\plong}\ntilde_\alpha(p)
( {~}^*\Gamma^{\mu\nu\alpha}(k,q,p)
+{}^*\Gamma^{\nu\mu\alpha}(k,q,p)) \cr
&~ -\gn2\softint ~{\Delta\alpha\over\plong p^2}
\left(2 {~}^*\Gamma^{\mu\nu}(k)-
  {}^*\Gamma^{\mu\nu}(k-p)-{}^*\Gamma^{\mu\nu}(q)\right) ~,\cr
&\,&(4.8)\cr}
$$
where the last line was obtained by applying the Ward identities
(3.1b) and (3.1c) successively. Note that $\Delta \beta(-p) = -
\Delta\beta(p)$ and $\Delta\alpha(-p)=\Delta\alpha(p)$.
\par
The gauge dependent contribution from the 3-point functions (2nd
term in (3.3)) is considerably more complicated. It can, however,
be simplified by applying the Ward identities whenever an
internal 4-momentum is contracted with one leg of either a three,
or two point function. The result can be summarized as follows:
$$\eqalignno{
&{Im}\left[<{}^*\Gamma^{(3)}~\Delta {}^*D {~}^*D
{~}^*\Gamma^{(3)}> +
<{}^*\Gamma^{(3)} {~}^*D~\Delta {}^*D {~}^*\Gamma^{(3)}>+
<{}^*\Gamma^{(3)}~\Delta {}^*D ~\Delta {}^*D {~}^*\Gamma^{(3)}>
\right]\cr
&= -\left\{X^\mu_\alpha(k) {~}^*\Gamma^{\alpha\nu}
+X^\nu_\alpha(k) {~}^*\Gamma^{\alpha\mu} + {~}^*\Gamma^{\mu\alpha}
Y_{\alpha\alpha'} {~}^*\Gamma^{\alpha'\nu} \right.\cr
&\,\,+g^2N\softint ~\left[ {\Delta\beta(p)\over \plong}
\ntilde_\beta(p)
{~}^*\Gamma^{\nu\beta\mu}(q,p,k)+(\mu\leftrightarrow\nu)\right]\cr
&\,\,+g^2N\softint ~{\Delta\alpha(p)\over\plong p^2}
{~}^*\Gamma^{\mu\nu}(q) \cr
&\,\,+g^2N\softint
\left[{\Delta\beta(p)\ntilde^\mu(p)\over \ntilde^2(p)}
\left({q^\nu\over q^2} -
{\beta_0(q)\ntilde^\nu(q)\over\ntilde^2(q)}
  \right)
+(\mu\leftrightarrow\nu) \right]\cr
&\,\, \left.
{-g^2N\over 2} \softint ~ {\Delta\beta(p)\Delta\beta(q)\over
\ntilde^2(p) \ntilde^2(q)}
\left[\ntilde^\mu(p)\ntilde^\nu(q) + (\mu\leftrightarrow\nu)
\right]
\right\} ~,
&(4.9)\cr}
$$
where we have used the identity:
$$
\ntilde_\mu(k) {~}^*\Gamma^{\mu\nu}(k)=\long \ntilde^\nu/\ntilde^2  ~,
\eqno(4.10)
$$
and defined:
$$\eqalignno{
X^{\mu\alpha}&=-g^2N\softint{\Delta\beta(q)\over\qlong}
\left(\ntilde_{\beta'}(q) {~}^*\Gamma^{{\alpha'}{\beta'}\mu}(p,q,k)
{~}^*{D'}_{\alpha'}^{\alpha}(p) +
 \ntilde^{\alpha}(q)\left[{p^\mu\over
p^2}-\beta_0(p)
{\ntilde^\mu(p)\over\ntilde^2(p)}\right]\right) \cr
&\,\,-g^2N\softint ~{\Delta\alpha(q)\over\qlong q^2}\left[
P^{\mu\alpha}(p)+\beta_0(p){p^\alpha\ntilde^\mu(p)\over
\ntilde^2(p)}\right]\cr
&\,\,-g^2N\softint ~{\Delta\alpha(q)\Delta\beta(p)\over\qlong q^2}
p^\alpha {\ntilde^\mu(p)\over\ntilde^2(p)}\cr
&\,\, -\gn2\softint ~{\Delta\beta(p)\Delta\beta(q)\over
\plong\qlong}
 \left(\ntilde_{\alpha'}(p)
\ntilde_{\beta'}(q) {~}^*\Gamma^{{\alpha'}{\beta'}
\mu}(p,q,k)p^\alpha\right)\cr
&\,\,+\gn2\softint ~\Delta\beta(p)\Delta\beta(q) \left[
{\ntilde^\mu(p)\ntilde^\alpha(q)\over\ntilde^2(p)\qlong} +
{\ntilde^\mu(q)\ntilde^\alpha(p)\over\ntilde^2(q)\plong}\right] ~,
&(4.11)\cr}
$$
and
$$\eqalignno{
Y_{\alpha\alpha'}&=
g^2N \softint~{\Delta\alpha(q)\over\qlong
q^2} {~}^*D'_{\alpha\alpha'}(p)\cr
&\,\,-g^2N \softint~ {\Delta\beta(p)\Delta\beta(q) \over
\plong\qlong}
\ntilde_\alpha(p)\ntilde_{\alpha'}(q)\cr
&\,\,+g^2N \softint~{\Delta\beta(p)\Delta\alpha(q)\over
\plong\qlong
q^2}
\left[\ntilde_\alpha(p) p_{\alpha'} + \ntilde_{\alpha'}(p)
p_\alpha\right]\cr
&\,\,+\gn2\softint~ {\Delta\alpha(p)\Delta\alpha(q) p_\alpha
p_{\alpha'}
\over\plong p^2\qlong q^2}    ~.
&(4.12)\cr}
$$
\par
Finally, the gauge dependent part of the ghost contribution is
$$\eqalignno{
{Im}<\Delta ghost>&= 2g^2N\softint  ~
{\Delta\beta(q)\over\ntilde^2(q)}
  \ntilde^{\nu}(q)\left( {p^\mu\over p^2} -\beta_0(p)
{\ntilde^\mu(p)\over\ntilde^2(p)}\right) \cr
&\,\,-g^2N\softint  ~ {\Delta \beta(p)\over\ntilde^2(p)}
{\Delta\beta(q)\over\ntilde^2(q)} \ntilde^\mu(p)\ntilde^\nu(q) ~.
&(4.13)\cr}
$$
It should be noted that in order to derive the above expressions,
extensive use was made of the identity (4.10) and:
$$
\softint f(p,q) = \softint f(q,p) ~,
\eqno(4.14)
$$
which is valid providing the cut-off is much greater than the
external momentum.
\par
After summing (4.8), (4.9) and (4.13), one finds a complete
cancellation of all terms that do not contain at least one factor
of the gauge invariant two point function
${~}^*\Gamma^{\mu\nu}(k)$. The final expression for the gauge
dependent
part of the imaginary part of the self energy, therefore takes
the simple
form:
$$
Im~\Delta\PPstar^{\mu\nu}(k)=Z(k) {~}^*\Gamma^{\mu\nu}(k)-
{~}^*\Gamma^{\mu\alpha}X_{\alpha}^{\nu}
-X^{\mu}_{\alpha} {~}^*\Gamma^{\alpha\nu}-
{~}^*\Gamma^{\mu\alpha}Y_{\alpha\alpha'} {~}^*\Gamma^{\alpha'\nu}  ~,
\eqno(4.15)
$$
where  we have defined
$$
Z(k)\equiv -g^2N\softint ~{\Delta\alpha(p)\over\plong p^2}    ~,
\eqno(4.16)
$$
and
$X^{\mu\alpha}$ and $Y^{\alpha\nu}$ are given in (4.11) and
(4.12) above. It should be stressed that Eq.(4.15) above
is valid for finite and/or infinitesmal gauge changes.
\par
Expression (4.15) contains the desired information about the
gauge dependence of the resummed self energy. First of all,
by contracting with $k_\mu k_\nu$, using the decomposition (3.7)
and the transversality of
${~}^*\Gamma^{\mu\nu}$, one can easily verify that the gauge
dependent part of
$\Pstar_d$ is identically zero to this order. This is consistent
with the Ward identity (3.16) which implies that, if
${~}^*\Pi_c\sim g^2T$, then
$\Pstar_d$ must vanish to this order.
\par
Secondly, by contracting $ Im ~\Delta\PPstar^{\mu\nu}$ in
(4.15) with
$k_\mu\ntilde_\nu$  one obtains:
$$
 Im~\Delta\PPstar_c= \long {1\over \ntilde^4 k^2} \ntilde_\alpha
X^{\alpha\nu} k_\nu                    ~.
\eqno(4.17)
$$
To derive (4.17) one needs to use the transversality of
${~}^*\Gamma^{\mu\nu}$, and the identity (4.10).
Since Braaten and Pisarski verified that $Im\Pstar_c$ vanishes on
the longitudinal mass shell in strict Coulomb gauge, we have
proven that this will
be true in any gauge, providing only that the coefficient of
$\long$ in (4.17) does not have a pole at $\long=0$. On
this basis one
can conclude that the non-transverse part of the
resummed self energy  will not contribute to the damping constant
of the longitudinal mode in any gauge.
\par
By contracting (4.15) with the projector $A_{\mu\nu}$,
one
finds that:
$$
Im ~\Delta\PPstar_t(k)= \tran Z(k) - \tran A_{\mu\nu} X^{\mu\nu} -
{(\tran)^2\over2}
A_{\mu\nu}Y^{\mu\nu}                   ~,
\eqno(4.18)
$$
while contracting with $\ntilde_\mu\ntilde_\nu$ yields
$$
Im~\Delta\PPstar_l(k)=\long Z(k)- 2 \long
{\ntilde_{\mu}\ntilde_{\nu}X^{\mu\nu}\over \ntilde^2}
    -{(\long)^2 \ntilde_{\mu}\ntilde_{\nu}Y^{\mu\nu}
 \over \ntilde^4}          ~.
\eqno(4.19)
$$
Thus, the gauge variation of the structure functions
vanishes on shell, providing that the corresponding coefficients
in the above equations do not develop poles on the mass
shell. As discussed in the introduction and in
Ref.[10], this implies the
gauge fixing independence of the resummed damping
constants, and generalizes the formal proof of Braaten and
Pisarski\ref{8}, who considered only Coulomb and Covariant
gauges.
In the next section we examine in more detail the integrals that
appear in (4.18).
\eject
\noindent
{\bf 5.~~~DAMPING RATE GAUGE DEPENDENCE FOR COVARIANT}

\noindent
{\bf~~~~~ AND COULOMB GAUGES}
\bigskip
Let us study the gauge variation of the transverse structure
function as an example. We will evaluate explicitly the
coefficients
of $k_T^2$ in (4.18) in order to check their behaviour on
mass-shell
($k_T^2 = 0$).
 \bigskip\noindent
{\bf Covariant Gauges}
 \par
 We start by investigating variations around a
fiducial covariant gauge, taking

$$ \Delta \beta =0~~~ (\beta = \beta_0 = 0)~,
 ~~~~ \Delta\alpha = \xi {\vec k_L^2\over k^2}$$

\noindent
 (with a trivial change of
notation for the gauge parameter as compared with (4.7)).

Using the expressions for $X^{\mu\nu}$ and $Y^{\mu\nu}$ as given
in
(4.11) and (4.12) and the following identities:
$$\eqalign{A_{\mu\nu} (k)~P^{\mu\nu} (p) &= 2 - {{{\vec p}^{~2}}
\over p^2}
+ {{({\vec k} \cdot {\vec p} )^2}\over{{{\vec k}^2} p^2}} \cr
A_{\mu\nu} (k)~A^{\mu\nu} (p)
&= 1 + {{({\vec k} \cdot {\vec p} )^2}\over
{{{\vec k}^2} {{\vec p}^{~2}}}} \cr
A_{\mu\nu} (k)~\tilde n^\mu (p) \tilde n^\nu (p)
&= {{p_0^2~ {\vec p}^{~2}}
\over p^4} \left( 1 - {{(\vec k \cdot \vec p )^2}\over {{\vec
k}^2
{\vec p}^{~2}}} \right)\cr
A_{\mu\nu} (k)~p^\mu p^\nu &= \vec p^{~2} - {{(\vec k \cdot \vec
p )^2}
\over {\vec k}^2} ~,\cr} \eqno(5.1)$$
where $p^2 = p_0^2   + {\vec p}^{~2}$,     we get
$$\eqalign{Im~ \Delta {^*\Pi_t} =
&- \xi g^2 N ~Im \Biggl[ - k_T^2 \int\limits_{{\rm soft}}
 {{dp}\over q^4}
\left( 1 -
{\vec p^{~2}\over p^2} + {{( \vec k \cdot \vec p )^2}\over {\vec
k^2
p^2}} \right) \cr
&+ {1 \over 2} k_T^4 \int\limits_{{\rm soft}}
 {{dp}\over q^4} \left\{ {1\over p_T^2} (1 +
{{(\vec k \cdot \vec p )^2}\over {\vec k^2 \vec p^{~2}}} ) +
{{p_0^2 \vec p^{~2}}\over {\vec p_L^{~2} p^4}}
 \left( 1 - {{ (\vec k \cdot
\vec p )^2}\over {\vec k^2 \vec p^{~2}}} \right) \right\} \cr
&+ {\xi\over 4} k^4_T \int\limits_{{\rm soft}}
 {{dp}\over {q^4 p^4}} \vec p^
{~2} \left( 1 - {{(\vec k \cdot \vec p )^2}\over {\vec k^2 \vec
p^{~2}}}
\right) \Biggr] ~.
\cr}\eqno(5.2)$$
It should be noted that, in the derivation of (5.2) from (4.18),
the analytic continuation of the external momentum is
implicitly assumed so that
 $Im~\Delta {^*\Pi_t} (k) \equiv Im~\Delta {^*\Pi_t}
(K_0, \vec k )$ .
\bigskip
In the following, we  evaluate the two integrals in (5.2) that
are potentially dangerous on-shell: $I_1 (k)$ and $I_2 (k)$
denoted
respectively as "transverse" and "longitudinal" and given by:
$$I_1 (k) = Im~\int\limits_{{\rm soft}}
 {{dp}\over {q^4 p_T^2}} ~,\eqno(5.3a)$$
and
$$I_2 (k) = Im~ \int\limits_{{\rm soft}} {{ dp~p_0^2~\vec
p^{~2}}\over
{q^4 p^4 \vec p^{~2}_L}}~.\eqno(5.3b)$$

The integrals in (5.2) which do not depend on the plasma
excitation properties $p_T$ and $\vec p_L$, respectively,
are treated in Ref.[19]; indeed they are well behaved on the
physical mass shell, $k_T = 0$.

\medskip\noindent
(i) \underbar{Calculation of the transverse integral}
\medskip
We restrict ourselves to the calculation
 for the gluon excitation at rest:
$\vec k = 0$. In order to evaluate the double pole we
use the identity
$${1\over q^4} = \lim_{m\rightarrow 0}
 \left( - {\partial \over {\partial
m^2}} \right) {1\over {q^2 + m^2}}~.\eqno(5.4)$$
\par
It is easy to derive that on the mass shell, i. e. when $K_0 =
m_g$,
the integral $I_1$ has no imaginary part except for $m=0$.
Therefore
we define $I_1 (m_g , \vec 0 )$ by taking the following limits:
$$I_1 (m_g , \vec 0 ) = \lim_{K_0 \rightarrow m_g}
\lim_{m\rightarrow 0}
\left\{ \left(- {\partial\over {\partial m^2}} \right) Im~
\int\limits_{{\rm soft}}
{{dp}\over {(q^2 + m^2 ) p_T^2}} \right\} ~, \eqno(5.5)$$
with $p=( w, \vec p ),~q = ( w^\prime, -\vec p ),~w^\prime = K_0 -
w$.
\medskip
Following the method of Refs.[8, 20], which we already used in a
related study of the fermion damping rate [11], we calculate the
imaginary part above by using the spectral representation of the
effective transverse gluon propagator
 $\Delta_t (p) = {1\over p_T^2}$  ~ [14, 21]
and of the free one $\Delta (q) = {1\over{q^2 + m^2}}$, as
$$\eqalign{\hat I \equiv
& Im~\int\limits_{{\rm soft}} {{dp}\over {(q^2 + m^2 ) p^2_T}}
\cr
 =  & {\pi\over
{n_B (K_0)}} \int {{d^3 p}\over {( 2\pi )^3}}
 \int\limits^{+\infty}_{-
\infty} d w ~d w^\prime~\delta (K_0 - w - w^\prime)
 n_B (w ) n_B (w^\prime ) \rho_t (w, p)
\epsilon (w^\prime ) \delta (w^{\prime 2} - E^2
),\cr}\eqno(5.6)$$
with $E^2 = p^2 + m^2$ and $\rho_t (w, p)$, the spectral
density associated to $\Delta_t (p)$ given in Ref. [21]. The
spectral density associated to $\Delta (q)$ is $\epsilon (w^
\prime ) \delta (w ^    {\prime 2} - E^2 )$,
and $n_B$ is the Bose distribution.
\par
Integrating with respect to $w^\prime$ and using the symmetry
property:

 $$\rho_t ( - w, p) =  -   \rho_t (w, p) ,$$

\noindent
 we get
$$\eqalign{\hat I =
& {1\over {8 \pi n_B (K_0)}} \int^\infty_0 {{pdp^2}\over E}
\int^\infty_0 d w \rho_t (w , p) \cr
&  \cdot \{ n_B (E) [ n_B (w) \delta (K_0 - w - E ) + (1 +
n_B (w)) \delta (K_0 + w - E )] \cr
&+ (1 + n_B (E)) [n_B (w) \delta (K_0 - w + E ) +
(1 + n_B (w)) \delta (K_0 + w + E )]\}.\cr}\eqno(5.7)$$
We first concentrate on the time like pole contribution in
$\rho_t (w,p)$:
$$\rho_t^{{\rm pole}} = \rho_t^{{\rm res}} (w, p)\delta (w -
w_t (p)), \eqno(5.8)$$
where $w_t (p)$ is the transverse mode dispersion relation
and is given in Refs.[14, 21], together with
$ \rho_t^{{\rm res}}$.
 Taking $K_0 > m + m_g$,
without loss of generality, we
obtain
$$\eqalign{\hat I^{{\rm pole}} =
&{1\over {8 \pi n_B (K_0)}} \Biggl[ {p_0\over E_0}
 \rho_t^{{\rm res}} (w_t
(p_0), p_0 ) n_B (E_0) \cr
& \cdot n_B (w_t (p_0))
 {1\over { \Biggl[{{\partial E}  /  {\partial p^2}} +
{{\partial w_t}  /  {\partial p^2}} \Biggr]}}_{p=p_0}
\Biggr],\cr}
\eqno (5.9)$$
with $p_0$ given by the $\delta$-constraint:
$$w_t (p_0) = K_0 - E_0 \equiv K_0 - \sqrt{p_0^2 + m^2} ~.
\eqno(5.10)$$
In order to perform the interesting limit $m\rightarrow 0$
 and $K_0 \rightarrow
m_g$, we notice that the solution of (5.10) vanishes for $m=0$ as
$$p_0 = K_0 - m_g + 0 (p_0^2 ),$$
using
$$w_t (p_0) {\strut\displaystyle\simeq \atop{p_0 \rightarrow 0}}
{}~~m_g + {3\over 5} {p_0^2
\over {m_g}} \pm .... \eqno(5.11)$$
Working out the derivative with respect to $m^2$ and taking $m
\rightarrow 0$, we find that the most singular term in the limit
$p_0 \rightarrow 0$ comes from the derivative

 $${\partial\over{\partial
m^2}} ({p_0\over E_0}) {\strut
\displaystyle
\longrightarrow \atop\displaystyle{p_0\rightarrow 0}}
{}~~- {1\over {2p_0^2}}~.$$

\noindent
 This allows to derive in the
appropriate limit:
$$I_1^{{\rm pole}} (m_g , \vec 0) {\strut\displaystyle
\simeq \atop{K_0\rightarrow m_g}}
{1\over {8\pi}} {T\over {2m_g}} {1\over{(K_0 -
m_g)^2}} ~.\eqno(5.12)$$
In order to conclude about $I_1$, we should now take care
about the contribution coming from the branch cut of the
effective
gluon propagator associated to Landau damping. In this case, from
first inspection, it seems safe to work directly on-shell and to
calculate
$$I_1^{{\rm disc}} (m_g , \vec 0 ) =
 \lim_{m\rightarrow 0} \left( -
{\partial \over {\partial m^2}} \right)
 \hat I^{{\rm disc}}~.\eqno(5.13)$$
Going back to Eq.(5.7), we end up by writing $I_1^{{\rm disc}}$
as
the sum of two terms $I^a$ and $I^b$,
$$\eqalign{& I^a \propto \lim_{m\rightarrow 0}
 \left( - {\partial\over
{\partial m^2}} \right)
 \int^{{\sqrt{m_g^2 - m^2}}}_{{{m_g^2 - m^2}\over {2mg}}}
p^2 dp {{n_B (E)}\over E} n_B (m_g - E) \rho_t^{{\rm disc}} (m_g
- E ,
p),\cr
& I^b \propto \lim_{m\rightarrow 0} \left( - {\partial\over
{\partial m^2}}\right)
\int^\infty_{{\sqrt{m_g^2 - m^2}}}~ p^2 dp {{n_B (E)}\over E}
(1 + n_B (E - m_g )) \rho_t^{{\rm disc}} (E- m_g , p),
\cr}\eqno(5.14)$$
neglecting irrelevant numerical factors and with $\rho_t^{{\rm
disc}}$,
the transverse spectral density associated to Landau damping,
given in
[21].
\medskip\noindent
When taking the deriative ${\partial\over{\partial m^2}}$  each
term yields two contributions:
\medskip
\item{-} the first one is given by the derivative of the
integration
bounds. In this case, the only source of potential singular
behaviour is
associated, in both $I^a$ and $I^b$, to the value of the
integrand
at $p = E=m_g$. Since $\lim\limits_{w\rightarrow 0} \rho_t^{{\rm
disc}} (w, p) \cdot n_B (w)$ is finite, no singularity is
encountered.
\medskip
\item{-} the second one is obtained when taking the derivative of
the
integrand. Writing ${\partial\over{\partial m^2}} = {1\over {2p}}
{\partial\over{\partial E}}$ and integrating by parts, we end up
as
above with a finite contribution.
\medskip\noindent
As a consequence, the contribution of $I_1 (k)$ to
 $Im~\Delta {^*\Pi_t} (k)$
is, in the appropriate limit - neglecting numerical factors,
$$Im~\Delta {^*\Pi_t}^{(I_1)} \propto \xi g^2 N
 {T\over {m_g}} { {k_T^4} \over {(K_0 - m_g)^2} }~ .
\eqno(5.15)$$
\par
With
$$k_T^2 = m_g^2 - {K_0}^2 {\strut\displaystyle\sim \atop
{K_0\rightarrow m_g}}
2m_g (m_g -K_0)~, $$
this leads to
$$Im~\Delta {^*\Pi_t}^{(I_1)} \propto \xi g^2 N T m_g  ~
.\eqno(5.16)$$
\bigskip\noindent
(ii) \underbar{Calculation of the longitudinal integral}
\medskip
In order to complete the argument, we need to calculate $I_2$.
This
in fact, using $p^2_0 = p^2 - \vec p^{~2}$, amounts to
calculating
two integrals:
$$I_{2a} = Im~\int\limits_{{\rm soft}} {{dp}\over q^4} {\vec
p^{~2}\over{
p^2 \vec p^{~2}_L}} ~,
\eqno(5.17)$$
and
$$I_{2b} = Im~\int\limits_{{\rm soft}} {{dp}\over q^4} {\vec
p^{~4}\over{
p^4 \vec p^{~2}_L}} ~.\eqno(5.18)$$
These two integrals may be calculated, following the same line as
above, from the basic integral
$$I (k) = Im~\int {{dp}\over {q^2 + m^2}} {1\over{p^2 + m^{\prime
2}}}
{1\over \vec p^{~2}_L}~ ,\eqno(5.19)$$
for $\vec k = 0$.
\noindent
This integral, analogous to a vertex integral, is evaluated with
the technique of Ref. [8], already used here which starts by
introducing "mixed" propagators for which an integral
representation in
terms of spectral densities is given. To be specific we introduce
the mixed propagator $\tilde\Delta (\tau, \vec q )$ as:
$$\Delta (q_0 , \vec q )= \int^{\beta = 1/T}_0 d\tau e^{iq_0\tau}
\tilde\Delta (\tau, \vec q)
= {1\over q^2+m^2}     ~,
\eqno(5.20)$$
and similarly for $\Delta (p_0 , \vec p)$ and
 $\Delta_L (p) = {1 \over {\vec p_L^2}}$.
\medskip
This allows to write $I(k)$ as
$$I(k) = Im~\int {{d^3 p}\over{(2\pi )^3}} \int^\beta_0 d\tau_1
\int^\beta_0 d\tau_2 e^{i\tau_1  k_0} \tilde\Delta (\tau_1
-\tau_2,
E_p) \tilde\Delta (\tau_1 , E_q )\tilde\Delta_L (\tau_2 , p) ~,
\eqno(5.21)$$
with $E_p = \sqrt{p^2+m^2}$, $E_q = \sqrt{p^2 + m^{\prime 2}} $,
and  in the following $p^2 = {\vec p}^{~2}$.
\medskip
The spectral representation which is then used in order to
calculate the imaginary part,
$$\tilde\Delta (\tau , E) = \int\limits^{+\infty}_{-\infty} d
w~e^{- w
\tau} \rho (w, E) (1 + n_B (w))\eqno(5.22)$$
is only valid when $0\leq \tau \leq \beta$ and therefore
$\tilde\Delta (\tau_1 -\tau_2 , E_p )$ should be defined to be
periodic with period $\beta$ [8]. This leads to (after analytic
continuation: $i k_0 \rightarrow K_0 + i \epsilon$)
$$\eqalign{    &  I (K_0 , \vec k = 0) =  \cr
&\pi
 (e^{\beta K_0}-1)
 \int {{d^3 p}\over {(2\pi )^3}}
 \int\limits
^{+\infty}
_{-\infty} d w_1~\int\limits^{+\infty}_{-\infty} d w_2
\int\limits^{+\infty}_{-\infty} dw
 {{\rho (w, E_p ) \rho (w_1 , E_q ) \rho_L (w_2 ,
p)}\over{w - w_2}} \cr
& \cdot  n_B (w_1 )  \{ n_B (w_2 ) \delta (K_0 - w_1 - w_2 )
   - n_B (w)\delta (K_0 - w_1 - w) \}.
\cr}\eqno(5.23)$$
The calculation then follows closely the steps described for
$I_1$.
Let us first consider $I_{2a}$ and focus on the pole contribution
in
the spectral function for the longitudinal mode $\rho_L (w, p)$:
$$\rho_L^{{\rm pole}} (w,p) =
\rho_L^{{\rm res}} (w,p) \delta (w - w_L(p)) =
 - {1\over p^2} {{w (
w^2 - p^2 )}\over {3m_g^2 - w^2 + p^2}} \delta (w -
w_L (p)), \eqno(5.24)$$
with $w_L (p)$ the dispersion relation given in Refs. [14, 21].
\par
The only interesting configuration in the limit $K_0
  \rightarrow m_g$ is
due to the term proportional to $\delta (K_0 - E_q - w_L (p))$ in
complete analogy with the case of the transverse integral. We
find
$$\eqalign{ I_{2a}^{{\rm pole}} (m_g , \vec 0)& = {1\over {8 \pi n_B
(K_0)}}
\left( - {\partial\over {\partial m^2}} \right)
\int\limits^\infty_0 {{pdp^2}\over E_q}
p^2 \int\limits^\infty_0 d w_2 \cr
& \cdot {{\rho_L^{{\rm res}} (w_2 , p )}\over{E^2_p - w_2^2}} n_B
(E_q)n_B (w_2) \delta (K_0 - E_q - w_2 ) ~.\cr}\eqno(5.25)$$
It turns out that there is a complete correspondence with the
calculation of $I_1$, i.e. replacing:
$$\rho_t^{{\rm res}} (w_t (p_0), p_0 ) \rightarrow
    {{p_0^2 \rho_L^{{\rm res}} (w_L (p_0),p_0)}\over{p_0^2 -
w_L^2 (p_0)}} ~,
\eqno(5.26)$$
 which both
equal ${1\over {2m_g}}$ when $p_0 \rightarrow 0$.
 We obtain therefore
the same singular value for $I_{2a}$ and $I_1$. For
$I_{2b}$, the additional differentiation with respect to
$m^{\prime 2}$
amounts to replacing the above correspondence by:
$$\rho_t^{{\rm res}} (w_t (p_0 ), p_0) \rightarrow
{{p_0^4 \rho_L^{{\rm res}} (w_L (p_0 ), p_0)}\over {(p_0^2 -
w_L^2 (p_0))^2}} ~,
\eqno(5.27)$$
which in the limit $p_0 \rightarrow 0$ yields $I_{2b} \simeq -
{p_0^2 \over {m_g^2}} I_1$. The integral $I_{2b}$ is therefore
negligible in the relevant on-shell limit.
\par
Going back to (5.2), we find that the gauge dependence of
the transverse structure function is indeed non vanishing on the
mass-shell
$$Im~\Delta {^*\Pi_t} (m_g , \vec 0 )=
 - \xi g^2 N k_T^4~ (I_1 (m_g ,\vec 0 )+
I_a (m_g , \vec 0 )) \propto g^2 N  T m_g ~, \eqno(5.28)$$
\noindent
which implies a gauge dependence for the damping rate (cf. Eq.(3.21)):
$$\delta\gamma_T \propto { {Im~\Delta {^*\Pi_t}}\over {m_g}}
\propto \xi g^2 N T ~,
\eqno(5.29)$$
for the transverse gluon excitation at rest.
\bigskip\noindent
{\bf  Coulomb gauges}
\bigskip
In this case  one may vary the effective propagator (Eq.(4.4)) as
$$^*D^{\mu\nu} \rightarrow~ ^*D^{\mu\nu} + \xi_c {{k^\mu
k^\nu}\over
{\vec k^4}}
\eqno(5.30)$$
around the strict Coulomb gauge $(\xi_c = 0)$. The gauge
dependent
part of the propagator can be written as in (4.6) with
$\Delta\beta
= 0$ and $\Delta\alpha = \xi_c {{\vec k_L^2 k^2}\over{\vec k^4}}$
so that the gauge variation $Im ~\Delta {^*\Pi_t}$
 is obtained from (5.2)
with trivial changes:
$$\eqalign{ Im~ \Delta {^*\Pi_t} = -\xi_c g^2 N~Im~
&\Biggl[ - k_T^2 \int_{{\rm soft}} {{dp}\over{\vec q{~^4}}}
\left( 1 - {\vec p{~^2}\over p^2}
+ {{(\vec k \cdot \vec p)^2}\over{\vec k^2 p^2}}\right) \cr
 + {1 \over 2} k_T^4 &  \int_{{\rm soft}} {{dp}\over{\vec
q{~^4}}}
 \left\{ {1\over p^2_T}
\left( 1 + {{(\vec k \cdot \vec p)^2}\over{\vec k^2 \vec
p{~^2}}}\right)
+ {{p^2_0 \vec p{~^2}}\over{\vec p^{~2}_L p^4}}
 \left( 1 - {{(\vec k\cdot
\vec p)^2}\over{\vec k^2 \vec p{~^2}}}\right)  \right\} \cr
&+ {\xi_c\over 4} k^4_T \int_{{\rm soft}} {{dp}\over{\vec q{~^4}
\vec p{~^2}}} \left( 1 - {{(\vec k \cdot \vec p )^2}\over{ \vec
k^2
\vec p{~^2}}} \right) \Biggr] ~.\cr}
  \eqno(5.31)$$
The only potentially dangerous contribution is proportional to
the integral
$$\int\limits_{{\rm soft}} {{dp}\over \vec q{~^4}}
{1\over{\vec p^{~2}_L p^4}} ~ . $$
 It does not depend, however, on $k_0$ and therefore
this yields -  after continuation - a vanishing imaginary part.
\medskip
We  thus conclude  that the problem of gauge dependence arises
in the general class of covariant
gauges but not in Coulomb gauges. The implications of our results
are
discussed in the next Section.
\eject
{\bf 6.~~~DISCUSSION}
\def\Ih{\hat I}
\medskip
In the previous Section we have shown by explicit calculations
that the
gluonic damping rate for the excitation at rest remains
gauge parameter dependent in the class of covariant gauges.
This extends the result previously derived for the fermionic rate
[11] in the same framework of the resummed effective perturbative
expansion [8].
\par
In the course of the calculation in
Section 5 it is stated that the integrals in Eq.(5.2)
 have no imaginary part on the
 mass-shell $K_0 = m_g$ for all values of the mass parameter
$m > 0$. This is most easily seen from the constraint (5.10)
which cannot be fulfilled when $w_t = m_g$; a solution is only
possible on the mass shell for $m = 0$.
One might therefore argue that continuity in $m$
requires the expression of Eq.(5.2) to vanish on shell for $m =
0$
as well; indeed this would be the case if the order of limits
(cf. Eq.(5.5)) were interchanged to
$
 \lim_{m\rightarrow 0}
 \lim_{K_0 \rightarrow m_g}  ~.
$
 However, the parameter $m$ was introduced as a
technical device to treat the
double pole $1/ (q^2)^2$ in
the gauge dependent term in Eq.(5.4), so the order of limits
is fixed as in Section 5.
Nevertheless this observation suggests that the observed
gauge dependence may be intimately related to infrared
divergences, despite the fact that the damping rate itself is
definitely
infrared finite.
\par
The discussion may be clarified
by introducing an infrared regulator, as advocated by Rebhan
[22] (see also refs.[23] and [24]).
In order to demonstrate  the problem, we
consider in detail the integral (cf. Eq.(5.6)):
$$ \Ih (k_0, {\vec k}    = 0) =
         ~\int {{dp}\over { q^2 ~ p^2_T}}~,
\eqno(6.1)$$
 which is an intermediate step in deriving a
 contribution to the self energy of the same
form as the first term in the second line of (5.2),
 namely
$$ I (k_0, {\vec k}    = 0) \sim
   k_T^4 ~\int {{dp}\over {(q^4)~ p^2_T}}~.
\eqno(6.2)$$
 The discussion
for these   integrals
may be generalized to the more complicated cases
corresponding to the integrals in Eqs.(5.3), (5.17) and (5.18),
and even to $\Delta {}^*\Pi_t$ (Eq.(5.2)) itself, before
taking the imaginary parts.
However,
in order to simplify the analysis,
 the  simpler dispersions [9]
$p^2_T = p_0^2 + {\vec p}^{~2} + m^2_g$, and
$\plong = \ntilde^2 p^2_T$ are used,
 instead of the physical
dispersion relations (Eqs.(3.9, 3.10)).
 \par
We start in the imaginary time-framework, $k_0 = 2 \pi m T$,
and later the continuation
$ik_0\to K_0+i\epsilon$
is performed.
In order to regularize possible infrared singularities
dimensional regularization is applied with analytically
continued spatial dimensions $n = 3 + 2 \epsilon$ [23]:
$$
\int~ dp \equiv T\sum_{p_0=2\pi m T}\int   {d^n p\over
(2\pi)^n}  ~.
\eqno(6.3)
$$
In the high temperature limit the leading term of (6.1) is:
$$\eqalign{
  \Ih (k_0, {\vec k}  = 0)  & \propto   {T \over {\sqrt {k_0^2}}
}~
\sqrt { { (k_0^2 + m^2_g)^2 }   \over {k_0^2} }
 \int_0^\infty ~ dp { {p^{2 \epsilon}} \over { p^2 +
       { {(k_0^2 + m^2_g)^2 }   \over { 4 k_0^2} } } }  \cr
    & \propto   {T \over {\sqrt {k_0^2}} }~
    \Gamma ( {1 \over 2} + \epsilon)
    \Gamma ( {1 \over 2} - \epsilon)
 ~\Bigl[ { {(k_0^2 + m^2_g)^2 }
 \over { 4 k_0^2} }   \Bigr]^{\epsilon}  ~. \cr}
\eqno(6.4)
$$
Irrelevant numerical factors are neglected.
As long as the energy $k_0$ is discrete and euclidean
there are no singularities present
and we are allowed to take $\epsilon = 0$.
In the next step the analytic continuation to the Minkowski
energy $K_0$ is performed, and the integral becomes imaginary:

$$   Im~\Ih (K_0, {\vec k}    = 0)
 \propto   {T \over  {K_0} }~ ,
\eqno(6.5)$$
and independent of $m_g$.
One may note that at non-zero temperature
$Im~\Ih$ is also defined for $K_0 < m_g$, in contrast to the
$T = 0$ contribution, which has only support for $K_0 > m_g$.

The integral
$  I (k_0, {\vec k}    = 0) $        of (6.2) contains
a term
$ \sim k^2_T~T
 \int_0^\infty ~ dp ~ {p^{2 \epsilon - 2}}$,
 which is infrared singular in three  spatial dimensions.
Therefore a non-vanishing regulator $\epsilon$ seems appropriate,
at least for the real part of the integral (after analytic
continuation).

However, it is remarkable that indeed these infrared singularities
cancel in $\Delta {}^*\Pi_t$ - at least in the described
approximation:
 $\Delta { }^*\Pi_t$
  has precisely the same structure as exemplified
by the integral in (6.4) !

The imaginary parts of the integrals (6.1)
 and (6.2) are smooth functions
in terms of the continued external energy;
for $\epsilon = 0$ and at the value
of $K_0 = m_g$ we find  e.g. for $\Ih$ from (6.4),
$$   Im~\Ih (m_g, {\vec k}    = 0)
 \propto   {T \over  {m_g} }~ ,
\eqno(6.6)$$
in agreement with the result for ${\hat I}^{pole}$ in Eq.(5.9)
for $w_t = m_g$ and $m=0$. I.e. a non-vanishing answer is
obtained with the following prescription:
first the necessary integrations are performed for discrete
external euclidean energies, where an infrared regulator may be
introduced, which is then removed before the analytic
continuation
to external Minkowski energies is done. The discontinuity
of the integrals has to be evaluated when staying off-shell, and
finally the on-shell limit is taken.
This prescription is equivalent to the calculation of the
imaginary
part of the integrals as described in Section 5;
it appears to be the appropriate procedure when dealing with the
effective euclidean
theory at finite temperature as it is implemented by
Braaten and Pisarski [8]. However,
it leads to a dependence on the choice of the
covariant gauge parameter.

The resolution of the gauge dependence problem advocated in
Refs.[22-24] follows the steps of taking the
on mass-shell limit just after the continuation, but keeping the
infrared regulator $\epsilon$ different from zero.
As emphasized by Rebhan[22], the correct treatment
of mass-shell singularities at zero temperature
requires an infrared regulator in the framework
of perturbation theory [18].
For the example above it is obvious (cf. Eq.(6.4))
that
$$ \Ih (K_0 = m_g, {\vec k}    = 0)   = 0   ~,
\eqno(6.7)$$
for $\epsilon > 0$, so there is no imaginary part
at $K_0 =m_g$, and gauge independence is restored!
\par
Although this prescription does appear to fix
the problem of gauge dependence at this order, it is perhaps
appropriate to make the following comments:
\item{1.}At zero temperature, the infrared regulator is required
to make the wave function renormalization well defined. In order
to evaluate the pole position, the regulator is not required.
In the present case, we are only interested in the pole position
and  in
the effective (truncated)
approximation under consideration for calculating damping
rates the necessity of the infrared regulator is not at all
obvious: only the imaginary parts of the integrals in (5.2) are
physically relevant (since higher loops can contribute to order
$g^2T$ to the real part of the self energy) and these imaginary
parts are not infrared divergent.
(We repeat that even the real part of
 $\Delta {}^*\Pi_t$ is free of infrared
singularities in the approximation investigated above.)
 Therefore
the limit $\epsilon\to0$ can be taken in (6.4) without
encountering a singularity.
Using the prescription in [22,23], however, a discontinuous
behaviour of the function
$Im ~ \Ih (K_0, \vec k = 0)$
(and therefore of $\Delta {}^*\Pi_t$)
 is enforced: it vanishes
at $K_0 = m_g$, whereas for $K_0 \ne m_g$ it behaves as
$T/K_0$ (in the limit that $\epsilon =0$).
In this way, the value of the damping rate, a supposedly physical
quantity, is affected by the presence, or absence of such a
regulator. Clearly, the question as to whether or not the
regulator is necessary in the expressions considered is
considerably more subtle than at zero temperature.
\item{2.}As long as $\Delta X$ in the gauge dependence identity
(1.1)
is well-behaved on mass-shell, gauge dependence is guaranteed
algebraically order by order in any self-consistent perturbative
expansion. In the presence of the mass shell singularities
discussed above, the gauge dependent terms in the lowest order
damping rate need to be ``regulated away". They do not vanish
algebraically, but instead rely on the presence of the regulator
to control the value of the integrals in question. This mechanism
for ensuring gauge independence is very different in this regard
from the algebraic proof, and it is not clear whether it will
work at higher orders. (This remains an important question, in
our opinion, even if such higher order contributions are
difficult, if not impossible, to calculate in practice.)
\item{3.}For gauge independence to hold, the infrared
regulator must preserve the Ward identities of the theory.
Dimensional regularization[23] does so trivially, but  results
in a somewhat strange analytic behaviour for the off-shell
propagator due to terms of the form $({K_0}^2-m_g^2)^\epsilon$. While
a cut-off avoids this particular problem[22], verifying the Ward
identities is somewhat more problematic.
\par
To summarize,
one appears to be left with the following choice: one can retain
an infrared regulator throughout the calculation, with potential
consequences as outline above, or
one may  use the
argument of continuity (this time in $K_0$) to follow the
prescription
advocated in Section 5 for calculating the damping rates in
the effective theory of Ref.[8].
However, if the latter choice is made, it remains an open problem
to determine whether or not there exist
missing terms[25] in the resummation which restore  (covariant) gauge
 independence of the damping rate.

 \par\vskip 15pt
\noindent{\bf ACKNOWLEDGEMENTS}\par
We kindly thank E.~Braaten, R.~Kobes, E.~Levin,
   R.~D.~Pisarski
         and  A.~Rebhan
 for helpful discussions.
 G.~K. is grateful to the Natural Sciences and Engineering
Research Council for support, and would also like to thank
LPTHE Orsay for its kind hospitality during the initial
stages of this work.
Partial support of this work by
 "Projets de Coop\'eration et d'Echange"
(PROCOPE) is gratefully acknowledged.
\vskip 1.00 truecm
\noindent{REFERENCES}\par
\baselineskip=15pt
\item{1)} $\,$ See for example, {\it The Proceedings of the
Workshop on Relativistic Heavy Ion Physics at Present and Future
Accelerators}, eds. T. Csorgo, S. Hegyi, B. Lukacs, and J. Zimanyi,
Central Research Institute for Physics Preprint KFKI-1991-28/A
(1991).
\item{2)} $\,$ See, {\it The Proceedings of the First Workshop on
Thermal Field Theories}, Physica {\bf A158} (1989).
\item{3)} For a review and references see: R.D. Pisarski, Nucl.
Phys. {\bf A525} (1991) 175c;  and E. Braaten, Nucl. Phys.(Proc.
Suppl.) {\bf B23} (1991) 351.
\item{4)}See for example S. Nadkarni, in Ref. 2) above, pp. 226-
234.
\item{5)}U. Heinz, in Ref. 2) above, p. 189-191; T.H. Hansson
and
I. Zahed, Phys. Rev. Lett. {\bf 58} (1987) 2397; Nucl. Phys.
{\bf B292} (1987) 725; S. Catani and E. D'Emilio, Phys. Lett.
{\bf
B238} (1990) 373.
\item{6)} O.K. Kalashnikov and V.V. Klimov, Phys. Lett. {\bf
B88}
(1979) 329; Phys. Lett. {\bf B95} (1980) 234; Sov. J. Phys.
{\bf 31} (1980) 699;
D.J. Gross, R. D. Pisarski, and L.G. Yaffe, Rev. Mod. Phys. {\bf
53}  (1981) 43;
U. Heinz, K. Kajantie, and T. Toimela, Phys. Lett.
{\bf B183} (1987) 96; Ann. Phys. (N.Y.) {\bf 176} (1987) 218.
\item{7)} R. D. Pisarski, Nucl. Phys. {\bf B309} (1988) 476; in
Ref. 2) above, p. 246; Phys. Rev. Lett. {\bf 63} (1989) 1129.
\item{8)}E.~Braaten and R.~D.~Pisarski, Phys. Rev. Lett. {\bf
64} (1989) 1338; Nucl. Phys.  {\bf B337} (1990) 569; Nucl.
Phys. {\bf B339} (1990) 310; Phys. Rev. {\bf D42} (1990) 2156.
\item{9)} $\,$
U.~Kraemmer, M.~Kreuzer, A.~Rebhan, and H.~Schulz,
Lecture Notes in Phys.  {\bf 361} (1990) 285.
\item{10)} $\,$
R.~Kobes, G.~Kunstatter, and A.~Rebhan,
Phys. Rev. Lett. {\bf 64} (1990) 2992;
Nucl. Phys.  {\bf B355}   (1991) 1.
\item{11)} R. Baier, G. Kunstatter, and D. Schiff, ``High
Temperature Fermion Propagator - Resummation and Gauge Dependence
of the Damping Rate", Phys. Rev. D (Rapid Communications) (to
appear).
\item{12)}
For a review: N.~P.~Landsman and Ch.~G.~van Weert,
Phys. Rep. {\bf 145} (1987) 141;
J.~I.~Kapusta, {\sl Finite-Temperature Field Theory }
(Cambridge University Press,
Cambridge, 1989);
 M.~Le~Bellac,
Lectures given at the XXXth Schladming Winter School,
Schladming, Austria,
 March 1991,
Lect. Notes in Phys. {\bf 396} (1991) 275.
\item{13)} J. Frenkel and J.C. Taylor, Nucl. Phys. {\bf B334}
(1990) 199; J.C. Taylor and S.M.H. Wong, Nucl. Phys. {\bf B346}
 (1990) 115.
\item{14)} V.V. Klimov, Sov. J. Nucl. Phys. {\bf 33} (1981) 934;
H.A. Weldon, Phys. Rev. {\bf D26}  (1982) 2789.
\item{15)}E. Braaten and R.D. Pisarski,
 Phys. Rev. {\bf D45} (1992) R1827.
\item{16)}A. Fetter and J.D. Walecka, {\it Quantum Theory of Many
Particle Systems} (McGraw-Hill, New York 1971).
\item{17)} R. Kobes, G. Kunstatter, and K. Mak, Z. Phys. {\bf
C45} (1989) 129.
\item{18)} C. Itzyskon and J.-B. Zuber, {\it Quantum Field
Theory} (McGraw-Hill, New York 1980).
\item{19)}
H.-Th. Elze, U. Heinz, K. Kajantie, and T. Toimela,
Z. Phys.  {\bf C37} (1988) 305.
\item{20)} $\,$
E.~Braaten, R.~D.~Pisarski, and T.~C.~Yuan, Phys. Rev. Lett.
 {\bf 64} (1990) 2242;
E.~Braaten  and T.~C.~Yuan, Phys. Rev. Lett.
 {\bf 66} (1991) 2183.
\item{21)}
 R.~D.~Pisarski, Physica {\bf A158} (1989) 146;
Fermilab preprint Pub - 88/113-T (unpublished).
\item{22)}A. Rebhan, ``Comment on ``High
Temperature Fermion Propagator - Resummation and Gauge Dependence
of the Damping Rate"", CERN preprint, CERN-TH/6434/92 (1992).
\item{23)} E. Braaten, private communication; E. Braaten and R.D.
Pisarski, "Calculation of the Quark Damping Rate in Hot QCD",
 BNL report-NUHEP-TH-92/3.
\item{24)} H. Nakkagawa, A. Ni\'egawa, and B. Pire, ``On the Gauge
Dependence Problem of the Fermion Damping Rate in Hot Gauge
Theories",
 CPT Ecole Polytechnique preprint A156-0292 (1992).
\item{25)} V.V. Lebedev and A.V. Smilga, Ann. Phys. (N.Y.) {\bf
202}
(1990) 229; Phys. Lett. {\bf 253} (1991) 231;
Physica {\bf A181} (1992) 187.
\bye